# Is the free energy landscape informative about transition rates? Lessons from the kinetic Ising model

Daniel Sigg,[1,a)] Vincent Voelz,[2] and Vincenzo Carnevale [3,b)]

[1]*dPET, Spokane, WA 99223, USA*

[2]*Department of Chemistry, Temple University, Philadelphia, Pennsylvania 19122, USA*

[3]*Institute for Computational Molecular Science, College of Science and Technology, Temple University, Philadelphia, Pennsylvania 19122, USA*

An oft-used concept in modeling macromolecules is the free energy landscape, obtained by coarse-graining a vast number of microstates into a low-dimensional mesh of mesostates. If the landscape contains two or more local minima (macrostates), one can compute global rate constants provided the dynamics of the dividing barrier regions are known. Here we compared experimental rate constants between ordered states in a kinetic Ising model with rates calculated from a coarse-grained master equation derived from the microcanonical ensemble. The coarse-grained macroscopic rate constants were roughly 50% larger than experiment across a range of environmental constraints, suggesting a systematic impediment of configurational progress on the microscopic scale that is specific to the structure of the Ising model. The error in coarse-graining lay with the calculation of the diffusion coefficient rather than with the shape of the free energy landscape, as ensemble- and time-averaged estimates of the latter were indistinguishable. Fluctuation analysis in the form of Nyquist's theorem also failed to substantially improve the value of the effective diffusion coefficient, suggesting a failure of the fluctuation-dissipation theorem. This study's findings from the Ising model raises doubts over the validity of the free energy landscape approach in calculating absolute transition rates for more complex systems such as proteins.

[a)] Electronic mail: dansigg@gmail.com

[b)] Electronic mail: vincenzo.carnevale@temple.edu

## I. INTRODUCTION

Small system dynamics are dominated by fluctuations. Single molecule recordings of proteins and other biological macromolecules reveal discrete transitions between macroscopic states. One of many functionally significant protein fluctuation is the characteristic random switching (gating) between conductance states in ion channels, a class of membrane-spanning pore proteins involved in cellular homeostasis and signaling. Ion channel gating is temperature-dependent and is typically sensitive to environmental variables such as voltage and ligand concentration [1,2]. Molecular dynamics (MD) simulation is an effective tool for sampling molecular fluctuations and exploring the effects of environmental perturbations in complex phenomena like ion channel gating. However, in its most basic form, this computational approach is limited to exploring trajectories spanning relatively short time-scales.

Since its inception, MD has benefited from importance sampling methods that estimate properties of the canonical distribution function by drawing samples from conveniently chosen auxiliary distributions (see G. Biondini [3] and references cited therein). In practice, these techniques are used to speed up the exploration of the configurational space, which may contain inaccessible regions due to the presence of large free energy barriers. If the relevant configurations are made equally probable by introducing appropriate biasing terms in the Hamiltonian, then relatively short trajectories can provide reliable estimates of thermodynamic observables [3–10]. The art of "enhanced sampling" lies in designing unbiased estimators that use these auxiliary trajectories and choosing the appropriate the set of "relevant" configurations to optimally describe the equilibrium properties of a system.

A common approach is to focus on a system's response to external perturbations. In the specific case of voltage-gated ion channels, one applies a sudden change in transmembrane potential ($V$) in order to monitor relaxation toward a stationary distribution of macrostates [11–14]. In thermodynamic terms, $V$ is a generalized force, and its conjugate displacement—the "gating" charge $q$—provides information about the system response. In general, one or more conjugate variables $x$ may be the natural choice for collective variable to parametrize a system's free energy landscape $W(x)$.

In this paper we address the related question of kinetics: Is the coarse-graining used to describe the system's equilibrium properties optimal to analyze its dynamics as well? In other words: Does spatial Brownian motion on $W(x)$ provide a good phenomenological model of the system's kinetics? The appeal of this approach is clear: the position-dependent diffusion coefficient $D(x)$ to be used in the associated Langevin equation can be obtained from the same simulation yielding $W(x)$ and therefore a single simulation setup could be used to enhance the sampling, characterize equilibrium and analyze kinetics [15–17].

However, it is not a priori clear whether or not this procedure automatically selects the optimal $x$ in the sense of the Zwanzig projection operator [18–20].

Here we evaluate the accuracy of the free energy landscape approach in a lattice model whose thermodynamics are completely determined by the microcanonical variables $E$ and $q$, where $E$ is internal energy and $q$ is the conjugate displacement to the applied field $V$. $E$ and $q$ were chosen as collective variables for the purpose of calculating kinetics. By averaging microscopic rate constants over the ($E$, $q$) ensemble and constructing a coarse-grained master equation, we were able to calculate global transition rates for any set of conditions in the operationally more useful ($T$, $V$) ensemble, where $T$ is temperature. The calculated rate constants were compared to the experimental values obtained from direct observation of the system unbiased Monte Carlo simulations, and found to be about 50% faster. The reason for this discrepancy is conjectured to be a reduction in available trajectory pathways in the system process due to forbidden transitions between microstates.

## II. THEORY

### A. Electrical Ising model

Inspired by the behavior of voltage-gated ion channels, we chose as our system an electrical variant of the 2D magnetic Ising model. The internal energy of the traditional magnetic model has the well-known form:

$$E(\sigma_1,\ldots,\sigma_N) = -J\sum_{\langle i,j\rangle}\sigma_i\sigma_j - \mu H\sum_i \sigma_i \qquad (1)$$

where $\langle i, j\rangle$ refers to nearest-neighbor interactions and particle spins $\sigma_i$ have value ±1.

The electrical Ising model has a slightly different but analogous form:

$$E(e_1,\ldots,e_N) = -\varepsilon\sum_{\langle i,j\rangle}\left(2e_ie_j - e_i - e_j\right) - \delta qV\sum_i e_i\ , \qquad (2)$$

where particle states $e_i$ may be 0 or 1. Oppositely aligned neighbors increase $E$ by an interaction energy $\varepsilon$. A change in $e_i$ from 0 to 1 moves a small "gating" charge $\delta q$, thereby decreasing the field energy by $\delta qV$. We consider only the "square" Ising grid with an even-numbered length $L$, and a relatively small number of total particles $N = L^2$ corresponding to $L$ = {16, 18, 20, 22}. Absent an applied field $V$, the internal energy of the system increases in increments of $2\varepsilon$ from 0 to $2N\varepsilon$, assuming periodic boundary conditions. The order parameter $q$, which serves as a reaction coordinate, is the total charge displacement:

$$q = \delta q \sum_i e_i \,, \qquad (3)$$

whose value ranges from 0 to $N\delta q$. The electrical Ising model can be easily mapped into the magnetic model using the transformations: $\sigma_i \leftrightarrow 2e_i - 1$, $2J \leftrightarrow \varepsilon$, $2\mu \leftrightarrow \delta q$, and $H \leftrightarrow V$, where $J$ is the coupling energy, $\mu$ is the magnetic moment, and $H$ is the magnetic field. Therefore all the results discussed here are valid also for the magnetic Ising model.

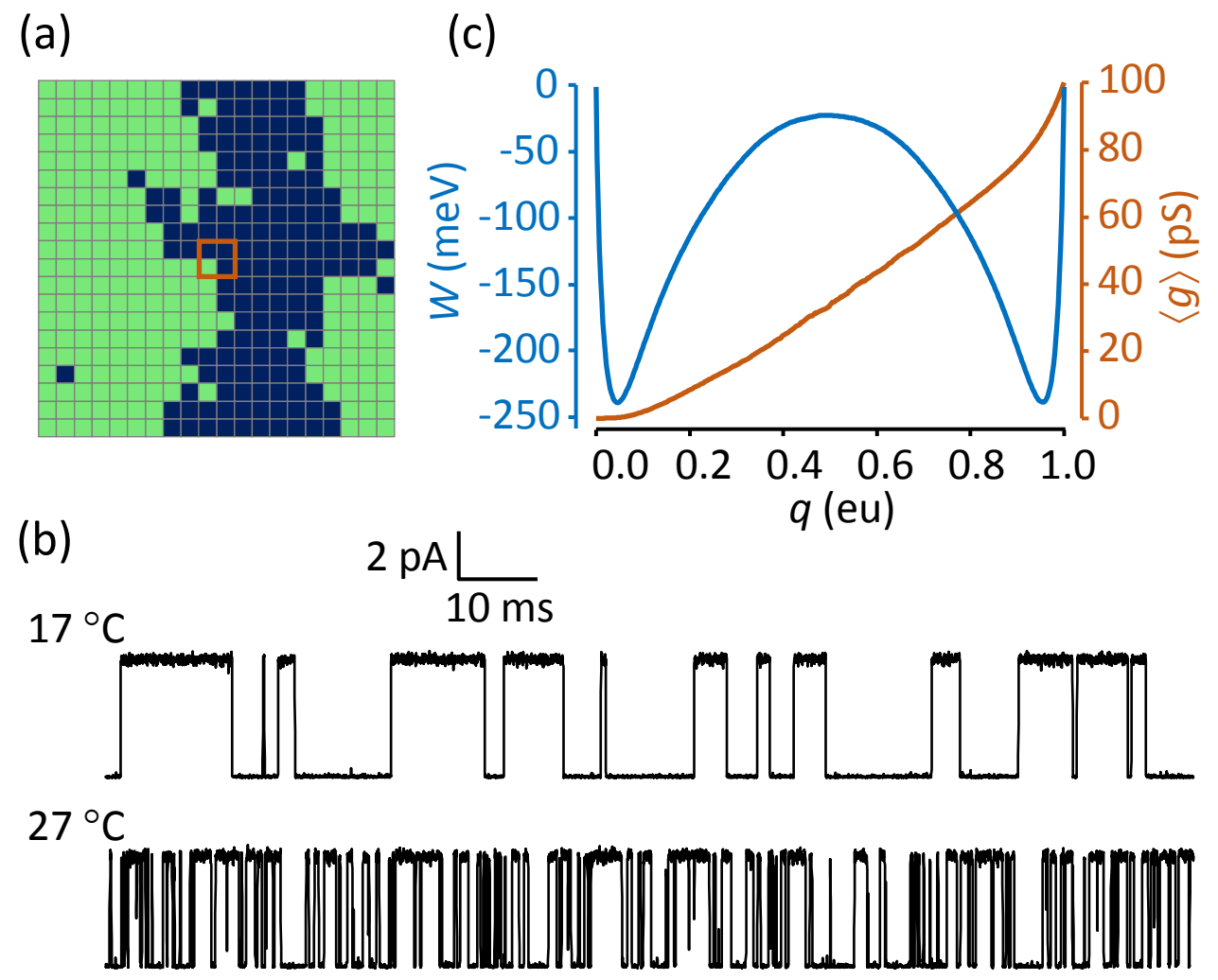

FIG. 1. (a) 20 x 20 electrical Ising model. Green cells indicate activated particles. In order to simulate an "ionic" current $i(t)$ that is qualitatively similar to those observed in ion channels, we devised a central "pore" consisting of a 2 x 2 particle sub-grid (brown square) whose instantaneous conductance $g$ is the sum of 25 pS (picosiemens) partial conductances of the four activated particles inscribed. (b) Monte Carlo simulations of the Ising model "pore" at two temperatures, with ionic current governed by Ohm's law: $i = g(V - V_{rev})$. Simulation parameters were as follows: $V = 0$ mV and $V_{rev}$ (reversal potential) = –60 mV. The current was digitally filtered as previously described [21] with a cutoff frequency $f_c = 10$ kHz and sampling rate $f_s = 100$ kHz. Observed noise was entirely due to internal fluctuations, with no external noise added. (c) 1D free energy landscape $W(q)$ and mean conductance $\langle g(q)\rangle$ obtained by time-averaging a 0.4 sec unfiltered trajectory at 17 °C. The abbreviation "eu" refers to the fundamental electric charge unit.

Simulated trajectories of the electrical Ising model with a centrally-conducting "pore" are qualitatively similar to laboratory-generated single-channel currents from ion channels (Fig. 1b). Both demonstrate stochastic switching between distinct macrostates, though through entirely different mechanisms. We made no attempt at a realistic depiction of channel gating; instead, the aim was to test whether slow barrier kinetics in a small system (the kinetic Ising model) were preserved after coarse-graining into a diffusion landscape. There are practical advantages to using the Ising model rather than performing detailed molecular dynamics simulations on a protein such as an ion channel: (1) There is no inertial memory in Ising dynamics, obviating the need for a time-dependent diffusion coefficient; (2) Particle interactions are nearest-neighbor and regular, allowing rapid calculation of kinetics and long-term trajectories; (3) The Ising system is bounded in $E$, $q$, and $N$, and its thermodynamics can be fully described using the microcanonical ensemble; (4) Microscopic rates can be expressed as functions of $E$ and $q$, and thus averaged over the microcanonical ensemble; (5) Coarse-grained kinetics are expressible in the form of a master equation that can be solved numerically.

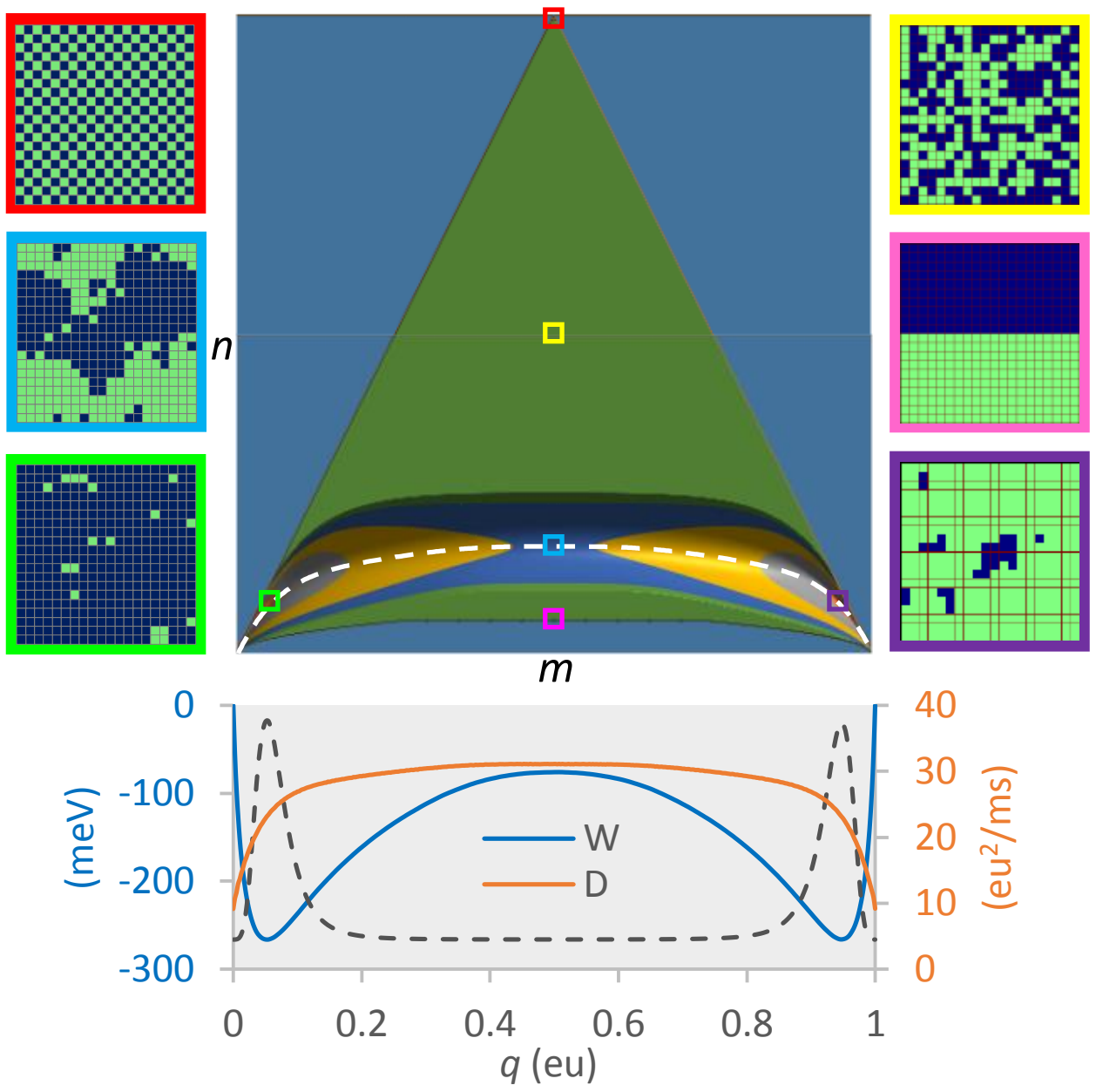

FIG. 2. Course graining the 20 x 20 electrical Ising model ($N$ = 400). The green triangle is the set of all microcanonical mesostates (= 73,448) on the ($m$, $n$) grid. The color-contoured region at the base of the triangle is the 2D reaction pathway comprised of a reduced number of mesostates (= 22,149) constrained by $W < W_{max}$, for conditions: $T$ = 22 °C, $V$ = 0 mV, and $W_{max}$ = 150 meV. The maximum vertical span (spatial bandwidth) of the reaction pathway is 68 cells, located at the saddle point. The dashed white line is the mean energy $\langle E \rangle$. Representative microscopic configurations are shown on both sides for: maximum energy (red); maximum entropy (yellow); saddle point (blue); minimum transition state energy (pink); and minimum free energy in state 1 (green) and state 2 (purple). Projected 1D landscapes $F(q)$ and $D(q)$ are plotted at bottom, along with the 1D equilibrium probability distribution $P(q)$ (dotted line).

The coarse-graining procedure for a 20 x 20 Ising model ($N$ = 400) is illustrated in Fig. 2. The microscopic dynamics are governed by a master equation, albeit an impossible one to solve due to the vast number ($2^N$) of microstates. We binned the microstates onto a more manageable $N$ x $N$ mesoscopic grid containing the microcanonical density of states $\Omega(q, E)$ [22,23]. Each mesostate ($q$, $E$) has a corresponding integer pair ($m$, $n$), where $q = m\delta q$ and $E = 2n\varepsilon$. The mesoscopic space of the Ising model is triangular with a concavity at the base. The height of the concavity is $2N^{½}\varepsilon$, giving rise to the well-known symmetry-breaking properties of the 2D Ising model. The purpose of coarse-graining was to construct a mesoscopic master equation whose smallest non-zero eigenvalue is the macroscopic relaxation rate.

Since realistic simulations are performed under constant $T$ and $V$, a free energy landscape $W(T, V)$ was constructed from $\Omega_{mn}$, representing the integer form of $\Omega(q, E)$.

$$W_{mn}(T,V) = 2n\varepsilon - kT\ln\Omega_{mn} - m\delta qV \ . \tag{4}$$

$W_{nm}$ is the potential of mean force (pmf) on the ($m$, $n$) grid. For sufficiently large $N$, it is effectively a smooth function of $q$ and $E$ consisting of two basins connected by a free energy barrier (Fig. 2). For positive values of $\varepsilon$, only the lower portion of the triangle is appreciably populated (corresponding to paramagnetism in the traditional Ising model). In order to limit the mesoscopic ($m$, $n$) states to a manageable number, we established a cut-off value $W_{max}$, chosen to be at least 6 $kT$ greater than the saddle point energy.

A further level of coarse-graining projected the 2D landscape $W_{nm}$ onto a 1D pmf $W_m$, shown at the bottom of Fig. 2. The projection is defined by:

$$\exp\left(-\frac{W_m}{kT}\right) = \sum_n \exp\left(-\frac{W_{mn}}{kT}\right). \tag{5}$$

The 1D pmf has $N + 1$ states. The 1D and 2D pmfs provide complete thermodynamic information, though quantities derived from the 1D landscape are limited to functions of $q$. From the 2D landscape we can, using Boltzmann averaging, obtain heat capacities for different ensembles (Fig. 3a); these are less peaked than for the infinite-particle Onsager solution around their respective "critical" temperatures. Both 1D and 2D pmfs yield the plot of mean charge displacement $\langle q\rangle$ versus $V$, which steepens around $V = 0$ at low temperatures (Fig. 3b).

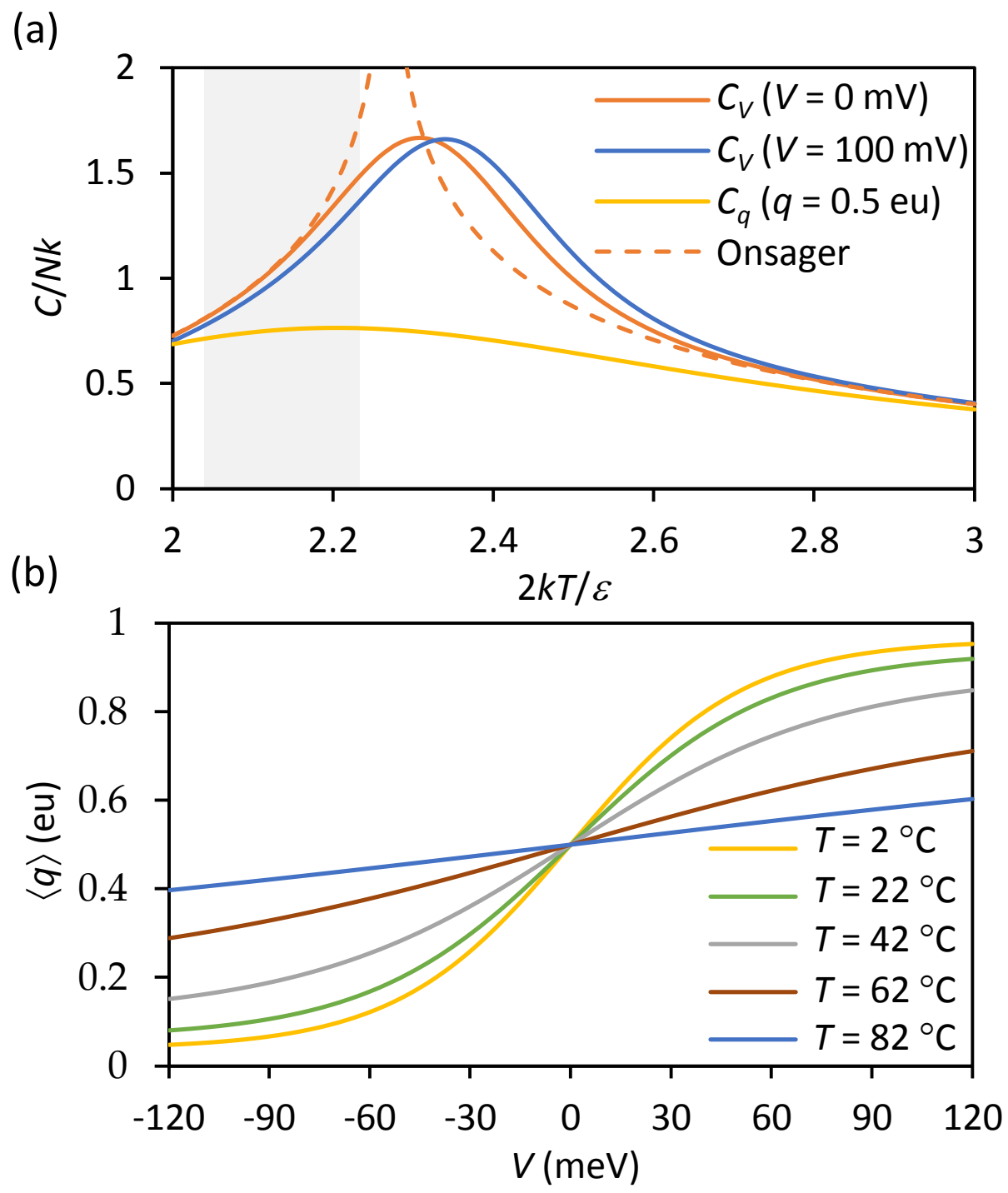

FIG. 3. Ising model thermodynamics. (a) Heat capacities $C$ of the 20 x 20 Ising model for constant $V$ and $q$, as a function of temperature. The dotted curve is the exact Onsager solution for infinite particles [24]. The shaded region denotes the physiological range of $T$ from 10 °C to 40 °C. (b) Mean charge displacement $\langle q \rangle$ versus $V$ for indicated temperatures.

## B. Microscopic rates

In order to study dynamics we needed a model for microscopic rates. These are constrained by detailed balance:

$$\frac{\alpha_i}{\beta_i} = \exp\left(-\frac{E(e_i = 1) - E(e_i = 0)}{kT}\right), \tag{6}$$

where $\alpha_i$ and $\beta_i$ are the forward and backward rates associated with the activation of the $i^{\text{th}}$ particle. We assumed activated barrier kinetics using the formulas:

$$\alpha_{u(i)} = \nu \exp\left(\frac{-\Delta\varepsilon_{u(i)}}{2kT}\right), \tag{7a}$$

$$\beta_{u(i)} = \nu \exp\left(\frac{\Delta\varepsilon_{u(i)}}{2kT}\right), \tag{7b}$$

where $\nu$ is a rate factor and the energy of activation $\Delta\varepsilon$ depends on the number $u(i)$ of activated neighbor particles, as follows:

$$\Delta\varepsilon_u(V) = 2(2-u)\varepsilon - \delta qV \ . \tag{8}$$

There is nothing unique about our choice of microscopic rate formulas. A different choice attributed to Glauber [25] could be used if one wishes to prevent microscopic rates from exceeding $\nu$, a useful property when performing kinetic Monte Carlo simulations with a fixed time step [26]. We employed Eq. 7 because activated processes are physically realistic, but they result in unbounded rate constants under the presence of large fields, resulting in inaccurate simulations of high frequency events if using a fixed time step. We avoid this problem through the use of the continuous-time Monte Carlo algorithm, as described in Numerical Methods.

## C. Mesoscopic rates

On the mesoscopic scale, there are ten cardinal transitions (five forward and five backward) available for a given $(m, n)$ state (Fig. 4), assigned to the possible combinations of $u$ (0…4) and $e$ (0, 1). The mesoscopic rate constants have the form: $a_{kk'} = \langle r_{kk'}\rangle\alpha_u$ (forward) and $b_{k'k} = \langle r_{k'k}\rangle\beta_u$ (backward), where $\langle r_{kk'}\rangle$ is the transition frequency from state $k = (m, n)$ to $k' = (m', n')$. Detailed balance is described by $\langle r\rangle_{kk'}\, g_k = \langle r\rangle_{k'k}\, g_{k'}$. Averaging the microscopic rates on the mesoscopic scale is the critical step in coarse-graining. For any square Ising model greater than $N$ = 4, the mesoscopic rate constants $a_{kk'}$ and $b_{k'k}$ are Boltzmann-weighted averages over transitions from configurationally distinct microstates. Not every cardinal direction is accessible to a particular microstate, even if other microstates within the same mesostate can undergo the move. We'll revisit this essential point in the Discussion section.

| Energy | $b$ coefficients | | $b_u$ | | $a_u$ | $a$ coefficients | |
|---|---|---|---|---|---|---|---|
| $2(n+2)\varepsilon$ | 4 | 0 | $b_4$ | | $a_0$ | 0 | 4 |
| $2(n+1)\varepsilon$ | 3 | 1 | $b_3$ | | $a_1$ | 1 | 3 |
| $2n\varepsilon$ | 2 | 2 | $b_2$ | $(m, n)$ | $a_2$ | 2 | 2 |
| $2(n-1)\varepsilon$ | 1 | 3 | $b_1$ | | $a_3$ | 3 | 1 |
| $2(n-2)\varepsilon$ | 0 | 4 | $b_0$ | | $a_4$ | 4 | 0 |
| | $4m - 2n$ | $2n$ | $(m-1)\Delta q$ | $m\Delta q$ | $(m+1)\Delta q$ | $2n$ | $4(N-m) - 2n$ |
| | Sum | | Charge | | | Sum | |

FIG. 4. Cardinal transitions from an ($m$, $n$) mesostate. Relationships between cardinal rates are expressed in the outer columns, where rate constants multiplied by $a$ and $b$ coefficients sum to the expressions on the bottom row. The sums express relationships between the cardinal rates. Detailed balance accounts for additional constraints between opposing rates, expressed as: $a_u/b_u = \exp(-\Delta\varepsilon_u/kT)$.

Projection of the 2D rate constants onto the $m$ axis through Boltzmann averaging yielded the 1D rates $a_m$ and $b_m$, or in continuous notation $a(q)$ and $b(q)$. In 1D, detailed balance assumes the form:

$$\frac{a(q)}{b(q+\delta q)} = \exp\left(\frac{W(q+\delta q)-W(q)}{kT}\right). \tag{9a}$$

The 1D diffusion coefficient $D(q)$ was assigned to the second coefficient of the Kramers-Moyal expansion of the coarse-grained master equation, yielding the following formula:

$$D(q) = \frac{1}{2}\left[a(q)+b(q)\right]\delta q^2 . \tag{9b}$$

The 1D kinetics can be expressed either through constructing the master equation from $a_m$ and $b_m$, or by inserting the landscape variables $W(q)$ and $D(q)$ into a Fokker-Planck or equivalent Langevin equation [27]. Outcomes in the form of global rate constants were similar for both representations since $a(q)$ and $b(q)$ are very slowly varying and nearly linear in the critical barrier region (data not shown). But since the diffusion landscape representation is at best a linearized approximation to the master equation, accurate only for vanishing $\delta q$ and noninteracting particles [28], we performed calculations exclusively using the master equation. Nevertheless, the diffusion landscape representation $\{W(q), D(q)\}$ remains very useful as an illustrative tool for understanding the stationary and dynamic properties of the system, so we will continue to make reference to it.

In addition to the aforementioned ensemble methods, we also employed time-averaging to determine the 1D variables $a_m$ and $b_m$, or equivalently $W(q)$ and $D(q)$. This consisted of unbiased simulation of the Ising model under equilibrium conditions of constant $T$ and $V$, from which time-averaged quantities such as the state probability $P(q)$ and transition rates $\langle a_m \rangle$ and $\langle b_m \rangle$ were obtained. For the 1D projection, we obtained $W(q)$ up to an additive constant by evaluating $-kT\ln P(q)$ over a very long trajectory. The boundary condition $W(0) = 0$ determined the additive constant. A faster method would have been to use umbrella sampling and align overlapping segments of $W(q)$ [29], but the advantage of a long unbiased run with numerous barrier crossings is it allowed us to also determine experimental transition rate constants from first passage times, as described in the next section (Macroscopic rates).

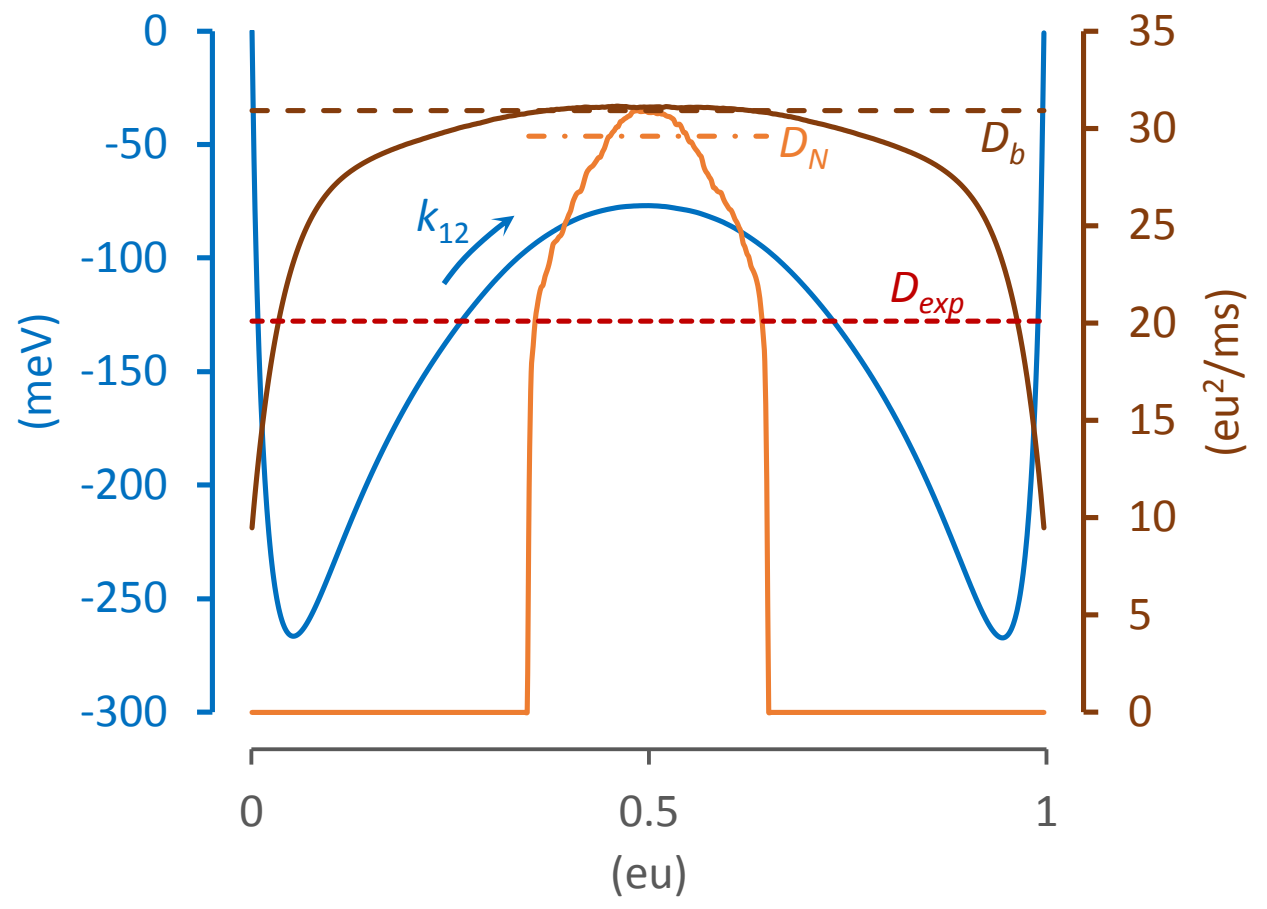

FIG. 5. Characterizing the diffusion coefficient. The diffusion landscape $D(q)$ obtained from the 1D master equation is shown as the solid brown line. Applying $D(q)$ to the free energy landscape $W(q)$ (blue line) in the form of a Smoluchowski equation [27] yields the forward rate constant $k_{12}$ = 0.921 kHz. The spatially independent value of $D$ (dashed brown line) predicting the same rate constant has a value (30.9 $eu^2$/ms), very close to the peak value of $D(q)$, which is $D_b$. High-bandwidth simulations performed with a sharp umbrella window in the barrier region ($q$ = 0.35 eu to 0.65 eu) yielded a truncated $D(q)$ (solid orange line; averaged from 500 traces). Nyquist's theorem applied to $i_g$ fluctuations measured from the windowed data set predicted $D_N$ = 29.1 $eu^2$/ms (dot-dashed orange line). By comparison, the constant value $D_{exp}$ consistent with the experimental rate constant (0.598 kHz) obtained from long-run Ising simulations is only 20.1 $eu^2$/ms (red dotted line). Model parameters were: $N$ = 400, $T$ = 22 °C, $V$ = 0 mV. The cut-off frequency for $i_g$ simulations using a digital Gaussian filter was: $f_c = 5 \times 10^5$ kHz. The filter bandwidth is given by $B = 1.064 f_c$ [30].

In our simulations, 1D ensemble and time-averaged quantities superimposed neatly, confirming ergodicity. A caveat to ensemble averaging is that it applies only to canonical variables that are functions of $q$ and $E$. Obtaining the mean and higher moments of a non-canonical variable, such as the "pore" conductance in Figure 1, can only be done through time-averaging. From a practical standpoint, the macroscopic rate is essentially determined by the maximum value of the diffusion coefficient at the barrier peak ($D_b$) (Fig. 5), which reduces the necessity of knowing the shape of $D(q)$ in the basin regions. This independence of the rate constant to the shape of $D(q)$ comes about if the free energy barrier is sufficiently large (≥ 6 $kT$) that basin states thermalize rapidly compared to the rate-limiting barrier flux.

Fluctuation analysis provided a third, independent method for determining $D$. From short (1 ms) segments of large-bandwidth trajectories confined to a 30% window centered on the $q$-axis, we measured fluctuations in the filtered impulses of charge movements $i_g = dq/dt$, known as "gating" current in the ion channel literature [27,30]. Applying Nyquist's current theorem: $4kTB/R = \langle i_g^2 \rangle$, where $R$ is resistance and $B$ is the recording bandwidth, the "Nyquist" value $D_N$ can be obtained using the Einstein relation $D_N R = kT$. We found the calculated value of $D_N$ in the barrier region to be nearly independent of filter bandwidth provided $f_c$ was much faster than the characteristic decay of the system. For example, in the reference model, an increase in $f_c$ from 1 x $10^5$ kHz to 5 x $10^5$ kHz resulted in only a 3% increase in $D_N$. The use of Nyquist's theorem in this context is analogous to other deployments of the fluctuation-dissipation theorem in obtaining the diffusion coefficient governing Brownian motion across a free energy landscape [16,17,31].

## D. Macroscopic rates

The third and final level of kinetic coarse-graining distills the kinetic Ising model to a single pair of global forward ($k_{12}$) and backward ($k_{21}$) rate constants connecting the macroscopic basin states. Starting with the 1D and 2D diffusion landscapes, we computed $k_{12}$ and $k_{21}$ using three methods: (1) eigenvalue analysis; (2) mean first passage time (MFPT); and (3) Monte Carlo simulation.

In the eigenvalue method, the mesoscopic rate constants were tabulated in a rate matrix **Q**. The upper diagonal terms of **Q** contained forward rates $\langle a \rangle$, and lower diagonal terms contained backward rates $\langle b \rangle$. Diagonal terms were assigned so that all row entries summed to zero, satisfying the master equation $d\mathbf{p}/dt = \mathbf{pQ}$, where $\mathbf{p}$(t) is the row vector of state probabilities [27]. The first non-zero eigenvalue of **Q** equals the sum $k_{12} + k_{21}$. In the absence of an applied field, we have $k_{12} = k_{21}$, and the problem of determining the global rates is solved, but even with $V \neq 0$, the ratio $k_{12}/k_{21}$ could be easily obtained from detailed balance.

The MFPT method involved truncating **Q** at $q = q_2$, where $q_2$ is the free energy minimum of basin state 2, thereby creating an invertible matrix $\mathbf{Q'}$. The MFPT for any $q$ to the left of the absorbing boundary at $q_2$ was obtained by solving $\mathbf{Q'\tau} = \mathbf{u}$ [28], where $\boldsymbol{\tau}$ is the column vector of MFPTs $\tau(q \rightarrow q_2)$ and $\mathbf{u}$ is the unit vector. The quantity $\tau^{-1}(q_1 \rightarrow q_2)$, where $q_1$ marks the minimum free energy in state 1, is an accurate estimate of the forward rate constant $k_{12}$ for sufficiently large barriers ($\geq$ 6 $kT$) [32].

In the last (Monte Carlo) method, we constructed a coarse-grained master equation for the 2D mesoscopic rates on the ($m$, $n$) grid, and one for the projected 1D rates $a_m$ and $b_m$, from which long (4 sec) Monte Carlo simulations generated first passage times between $q_1$ and $q_2$, the inverse averages of which yielded $k_{12}$ and $k_{21}$.

### E. Ising Monte Carlo simulation

A continuous-time kinetic Monte Carlo algorithm was used to obtain first passage times from $q_1$ to $q_2$ in the Ising model for constant $T$ and $V$. We obtained the inverse MFPT drawn from thousands of simulated transition events and used this for the experimental value of $k_{12}$ in the Ising model, which was compared to results from coarse-grained master equations. It was not possible to verify the accuracy of the simulations using independent methods, as we did for the coarse-grained models. However, consistent outcomes were obtained using several "good" pseudorandom number generators (Fig. 6), suggesting the MFPT values were reliable. Also, the measurement error obtained from multiple simulation experiments was consistent with the small standard deviation expected from the exponential distribution of first passage times.

## III. NUMERICAL METHODS

The majority of Monte Carlo simulations and analysis routines were run using customized software written in C. For each set of environmental conditions ($N$, $T$, $V$, $W_{max}$), six simulations ran simultaneously, comprising a single experiment. Each run generated the following: (1) the density of states $\Omega$ and transition frequencies $\langle r \rangle$; (2) 1D and 2D potentials of mean force and rate constants in the ($N$, $T$, $V$) ensemble; (3) simulated trajectories of the Ising model and coarse-grained master equation simulations in 1D and 2D; (4) MFPT for Ising, 1D, and 2D Monte Carlo simulations using first passage times from ~$10^4$ events, and for 1D and 2D master equations using the $\mathbf{Q'}$ method; and (5) eigenvalue analysis for 1D and 2D master equations. Output parameters for each experiment were expressed as mean and standard error. We chose as a reference model the 20 x 20 Ising grid ($N$ = 400) with the following parameters: $T$ = 22 °C; $V$ = 0; $\varepsilon$ = 24 meV; $q_{max}$ = 1 eu; and $\nu = 5 \times 10^4$ kHz. Additional experiments with different $T$, $V$, and $N$ were run to determine environmental and grid size effects on the forward rate constant. A total of 15 experiments were performed. The values of $\varepsilon$, $\delta q = q_{max}/N$, and $\nu$ were held constant across experiments. The value of $W_{max}$ was either 150 or 200 meV, depending on the proximity of the 2D saddle point to the zero-energy baseline.

### A. Random number generators

We tested several pseudorandom number generators to address the possibility of biased sampling. These were as follows: (1) Numerical Recipes© long-period L'Ecuyer generator with Bays-Durham shuffle (ran2) [33]; (2) Intel® MKL Library SIMD-oriented Fast Mersenne Twister (FMT) [34]; (3) Intel® MKL Library multiplicative congruential generator (MCG) [35]; and (4) Intel® MKL Library combined multiple recursive generator (MRG) [36]. Three of the four generators yielded similar results. The MCG algorithm exhibited greater variability and systemic biases compared to the other three (Fig. 6), and is not recommended for use in our model. Most simulations employed the ran2 algorithm. In a small number of cases involving large barriers where the performance of the random number generator was in doubt (for example, Fig. 6b), repeating the simulations using the FMT and MRG algorithms helped to confirm the results.

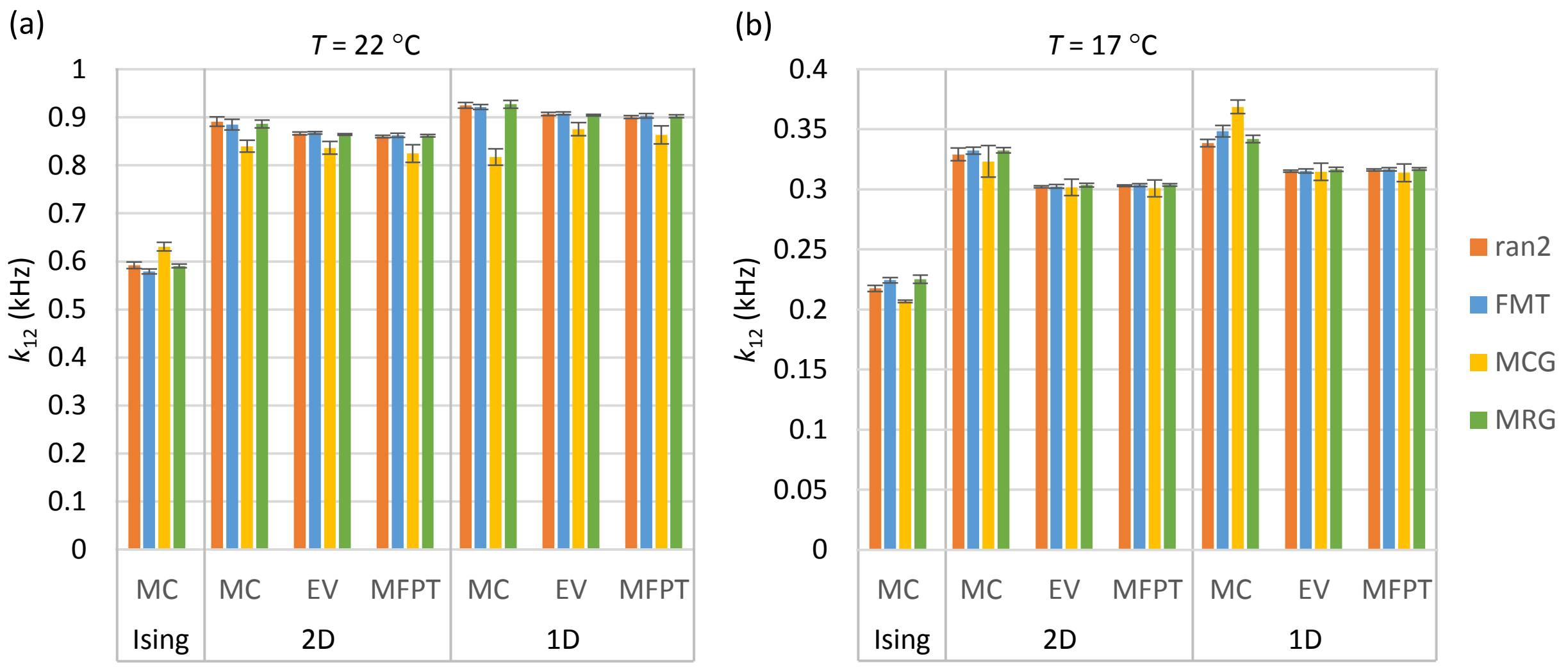

FIG. 6. Forward rate constant estimates (mean ± s.e., $n$ = 6) from different methods (MC = Monte Carlo, EV = eigenvalue analysis, MFPT = mean first passage time) as a function of pseudorandom number generator. (a) Reference model; $T$ = 22 °C. (b) Temperature lowered to 17 °C, which increased the barrier height and led to a discrepancy between Monte Carlo and the other two methods (EV and MFPT). Results were consistent between generators except in the case of the multiplicative congruential generator (MCG), which demonstrated lower accuracy and precision compared to the other three generators.

## B. Computing mesoscopic rates

The Metropolis Monte Carlo (MMC) method was used to compute average rates in the microcanonical distribution. Applying traditional MMC for infinite $T$ and zero $V$ allows every trial move to be accepted, in principle yielding the microcanonical ensemble. However, the large entropy gradient makes this algorithm impractical for all but the smallest systems, so a bias potential $\omega_k$ was applied to "flatten" the gradient and allow uniform sampling over all $k$ = ($m$, $n$) mesostates. The 1/t variant of the Wang-Landau algorithm with endpoint refinement parameter $F_{final} = 10^{-6}$ was used to rapidly obtain an estimate for $\ln\omega_k$ [37]. MMC was then implemented using the following procedure: pick a random particle for a trial flip, and after determining the locations of the current state $k$ and the trial state $k'$ on the ($m$, $n$) grid, apply the following acceptance criterion for the $k \rightarrow k'$ transition:

$$r_n \leq \frac{w_k}{w_{k'}}, \tag{10}$$

where $r_n$ is a uniform random number from 0 to 1. If Eq. 10 is satisfied, flip the particle. The time is incremented whether or not the trial was successful. For each discrete time point $s$, the ten cardinal rates $a_u$ and $b_u$ were determined by summing microscopic rates over all particle transitions. For example, if nine out of $N$ particles contributed to the backward transition $(m, n) \rightarrow (m - 1, n + 2)$, then the instantaneous rate $b_4(s)$ was assigned the value $9\beta_4$, where $\beta_4$ is the corresponding microscopic rate (Eq. 7b). The starting configuration for each simulation run was chosen randomly. Following an equilibration period of $2 \times 10^6$ time points, time-averaged forward rates were obtained over the next $2 \times 10^{10}$ time points using the following formula:

$$\left\langle a_{u(k)} \right\rangle = \frac{\sum_s \omega_k(s) a_{u(k)}(s)}{\sum_s \omega_k(s)} \alpha_{u(k)}, \tag{11}$$

with a similar expression for reverse rates $\langle b_{u(k)} \rangle$. The ratio of sums in Eq. 11 is the mean transition frequency $\langle r_{kk'} \rangle$, which can be rewritten: $\langle r_{u(k)} \rangle_a$, and for the reverse transition: $\langle r_{u(k)} \rangle_b$. The microcanonical density of states was obtained from:

$$\ln\Omega_k = N\ln(2) + \ln\frac{\sum_s \omega_k(s)}{\sum_{ks} \omega_k(s)}. \tag{12}$$

The density of states $\Omega_k$ and rate factors $\langle r_{u(k)} \rangle_a$ and $\langle r_{u(k)} \rangle_b$ were stored as $N \times N$ matrices $\ln\mathbf{\Omega}$, $\mathbf{A}_u$, and $\mathbf{B}_u$ (Fig. 7). These eleven matrices are correlated (Fig. 4) and could theoretically be stored as five independent matrices, but to avoid a possible increase in sampling error this was not done. It should be emphasized that, like $\ln\mathbf{\Omega}$, $\mathbf{A}_u$ and $\mathbf{B}_u$ are statistical matrices that depend only on $N$. In order to calculate mesoscopic rate constants as a function of environmental constraints: $T$, $V$; and model parameters: $\varepsilon$, $q_{max}$, and $\nu$, we used: $a_{u(k)} = \langle r_{u(k)} \rangle_a \alpha_u$ and $b_{u(k)} = \langle r_{u(k)} \rangle_b \beta_u$, employing Eqs. 6 and 7 for the microscopic rates $\alpha_u$ and $\beta_u$.

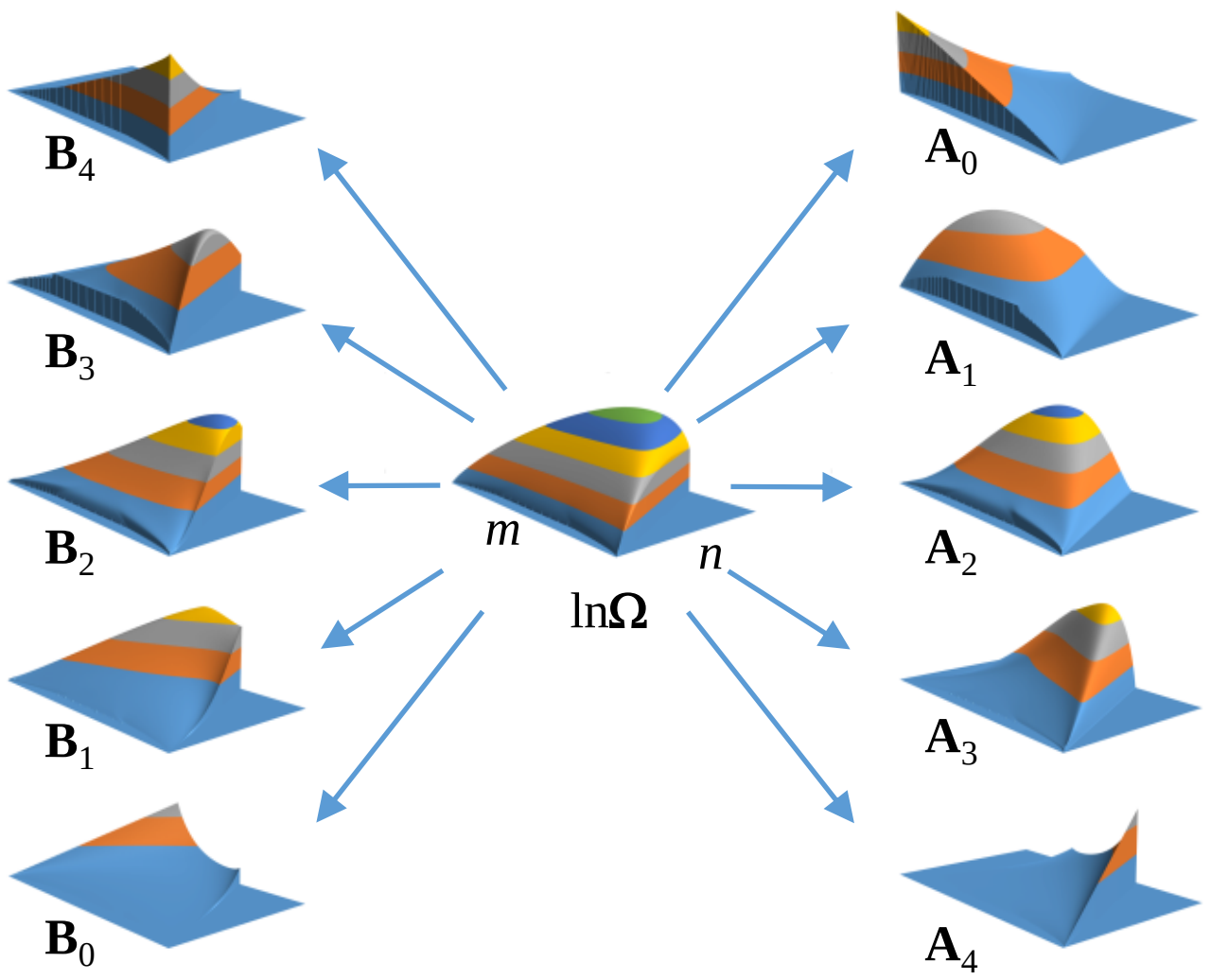

FIG. 7. Density of states (ln**Ω**) and cardinal rate (**A**, **B**) matrices for 20 x 20 Ising model. Vertical axes are not to scale. The *n*-axis was cut off at $n$ = 254 due to limitations of the plotting software.

### C. Computing macroscopic rates

Macroscopic rates $k_{12}$ and $k_{21}$ were calculated from the rate matrix **Q** using the methods (Monte Carlo, eigenvalue, and MFPT) detailed in the Theory section. A practical difficulty arose from the large size of **Q**, particularly for the 2D representations. From the $(N+1)^2$ possible 2D states, somewhat less than half are theoretically accessible (Fig 2). In the case of the 20 x 20 Ising model, the 2D size of **Q** would be 73,448 x 73,448. The number of accessible states $l$ can be substantially reduced by establishing a practical free energy cutoff $W_{max}$ chosen to be at least 150 mV higher than the saddle point energy. The danger in choosing too small a value for $W_{max}$ is that this narrows the accessible region of the barrier, thereby underestimating the calculated transition rate. Further size reductions in **Q** were achieved by repacking the matrix into band form. We numbered the mesostates in column order of increasing $m$ and $n(m)$, resulting in a maximum spatial bandwidth $d$ for the main diagonal that is equal to the width of the saddle point region. The banded **Q** dimensions were $l$ x $2d+1$. Typical values of $l$ and $d$ for the 2D landscape were ~20,000 and ~70. By taking advantage of eigenvalue and linear routines designed for band matrices, macroscopic rates were calculated in short order. The eigenvalue and mean first passage time calculations were easily adapted for 1D calculations, where **Q** is tridiagonal ($d$ = 1) and $l = N + 1$.

## D. Macroscopic rate constants from Monte Carlo simulation

A continuous-time kinetic Monte Carlo algorithm [27,38] was used to simulate kinetics of both the Ising model and coarse-grained representations. The continuous-time method avoids the shortcomings of the Metropolis algorithm, which is not suitable for kinetics, and of discrete-time kinetic Monte Carlo methods, which distorts high frequency events for insufficiently small time steps. In simulating the Ising model, which is governed by a master equation, the state-dependent flip rate ($\alpha_i$ or $\beta_i$) for every particle $i$ was determined at each step, and particle rates were summed in the forward ($\sum\alpha_i$) and backward ($\sum\beta_i$) directions. The real-valued time between transitions was randomly obtained from $-\ln r_n/r$, where $r$ is the combined transition rate $\sum\alpha_i + \sum\beta_i$, and the particle to be flipped was chosen with a separate random draw using weights proportional to each particle's activation or deactivation rate. The process was repeated until 4 seconds of simulation time was completed. A similar algorithm was used to simulate the coarse-grained master equations in 1D and 2D as described by **Q**.

# IV. RESULTS

## A. Global rate constants from Monte Carlo Simulation

First passage times from ~$10^4$ macroscopic transitions simulated from Ising dynamics were used to compute the experimental values of the global rate constants $k_{12}$ and $k_{21}$. We consider from this point only the forward rate $k_{12}$ since, apart from two experiments where $V \neq 0$, the opposing rates are equal. Filtered Monte Carlo trajectories on a time scale that includes multiple transition events clearly demonstrate the difference in rate constants between Ising and coarse-grained dynamics, visible by eye (Figs. 8a, 8b). Very precise values of $k_{12}$ could be obtained by calculating the mean dwell times within the first basin ($\tau_1$). We equated $\tau_1$ with the MFPT from $q_1$ to $q_2$. The two-state approximation predicts an exponential distribution of dwell times. Figure 8c shows the distributions from Ising, 1D, and 2D Monte Carlo simulations. A Sigworth-Sine transformation was applied to the dwell time distribution in order to generate uniform residuals for fitted curves [39]. A practical feature of the Sigworth-Sine plot is that the transformed distribution peaks at $\tau_1$.

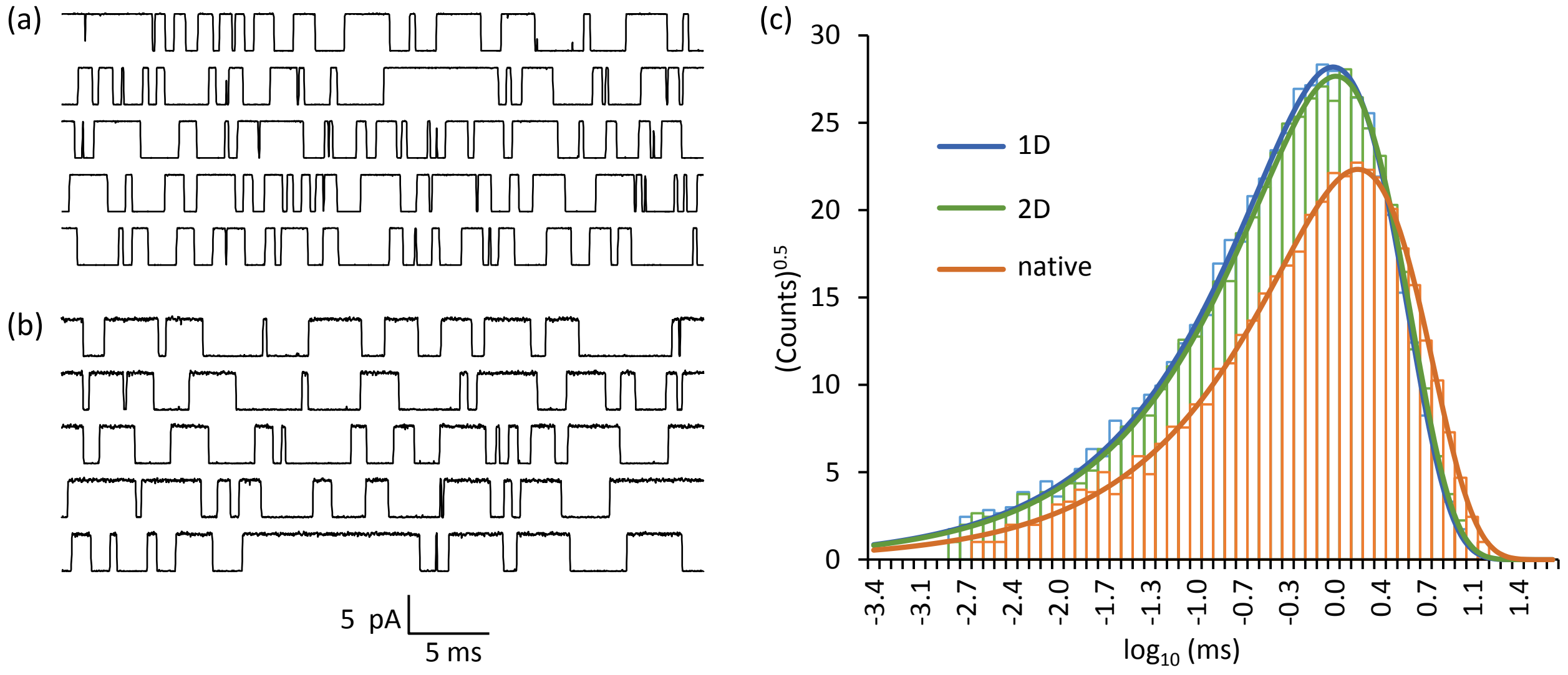

FIG. 8. Dwell time analysis. (a) Filtered "ionic" current trajectories simulated from the 1D master equation. (b) Same as (a) except with Ising dynamics. The reference model was used in both simulations; $f_c$ = 10 kHz. (c) Sigworth-Sine plot of binned passage times from $q_1$ to $q_2$ demonstrates fewer overall events and a longer average dwell time for Ising dynamics compared to the coarse-grained 1D and 2D dynamics.

It is apparent from Fig. 8 that for both the Ising and coarse-grained models, the binned dwell times faithfully followed the theoretical distribution across roughly four orders of magnitude in time. The macroscopic two-state approximation is evidently very good across relevant time scales. Even allowing for binning error, fitting the Sigworth-Sine plot to the dwell time distribution yielded a $\tau_{fit}$ that was very close to $\tau_1$ (Table I).

TABLE I. Forward Mean First Passage Times from Monte Carlo simulation.

| | Ising | 2D | 1D |
|---|---|---|---|
| counts | 6,776 | 10,397 | 10,791 |
| $\tau_1$ (ms) | 1.672 ± 0.020[a] | 1.145 ± 0.011[a] | 1.086 ± 0.010[a] |
| $k_{12}$ (kHz) | 0.5982 ± 0.0073[b] | 0.8735 ± 0.0085[b] | 0.9210 ± 0.0088[b] |
| $\tau_{fit}$ (ms) | 1.675 | 1.146 | 1.088 |

[a]standard error ($\tau_1$) = $\tau_1$/counts$^{1/2}$ (property of exponential distribution).
[b]standard error ($k_{12}$) = ($k_{12}/\tau_1$)s.e.($\tau_1$).

The coarse-grained representations (1D and 2D) predicted significantly smaller values of $\tau_1$ than the Ising model, yielding a roughly 50% larger $k_{12}$ value. The 2D result was marginally more accurate than the 1D rate. These are the major findings of our study.

## B. Comparing numerical methods

We checked the Monte Carlo estimate for $k_{12}$ against the eigenvalue and MFPT methods for the 1D and 2D representations. There was generally excellent agreement between the three calculations. There were some mild discrepancies in two circumstances where the PMF barrier exceeded 8 $kT$ ($T$ = 17 °C and $N$ = 484). The Monte Carlo-derived rate constants for these taller barriers were slightly larger compared to the other two methods (for example, Fig 6b). However, even in these special cases, the Monte Carlo outliers were not enough to significantly alter the mean value of $k_{12}$ taken from all three methods.

## C. Global rate constants as a function of N, T, and V

The dependence of $k_{12}$ on structural and environmental variables $N$, $T$, and $V$ can be predicted from the shapes of 1D energy and diffusion landscapes shown in Fig. 9. Reactive rate theory predicts that $k_{12}$ should decrease exponentially with forward barrier height $\Delta F$ and increases linearly with maximum $D$, with relative curvatures around both the reaction state (basin 1) and the barrier making minor contributions [40]. The value of $\Delta F$ increased with increasing $N$ and decreased with increasing $T$ and $V$. The barrier value of the diffusion coefficient ($D_b$) increased substantially with increasing $N$, more slowly with increasing $T$, and was negligibly dependent on $V$.

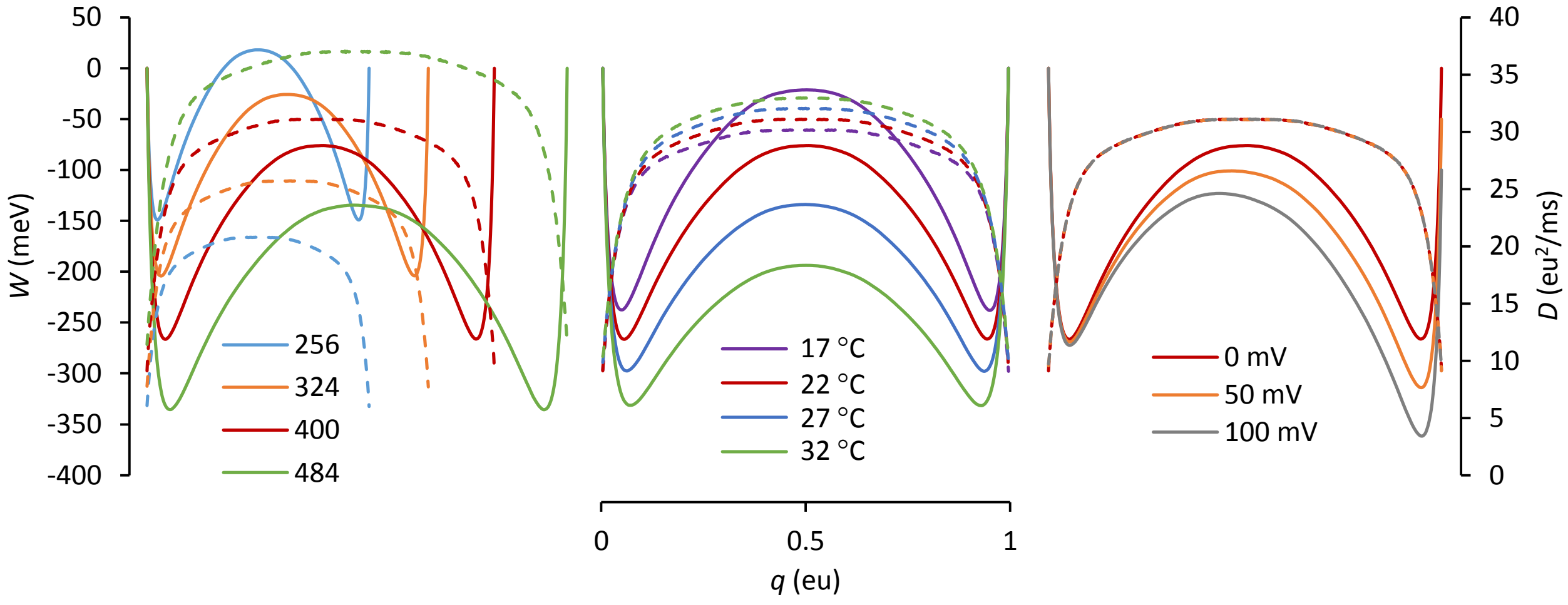

FIG. 9. Free energy (solid lines) and diffusion (dashed lines) landscapes as a function of model parameters (left plot, $N$; middle plot, $T$; right plot, $V$). The red curves correspond to the reference model ($N$ = 400; $T$ = 22 °C; $V$ = 0 mV). Landscapes were calculated from ensemble averaged mesoscopic rate constants, but time averaging of Monte Carlo trajectories yielded practically identical results.

We wanted to know if coarse-graining overestimated the macroscopic rate constant consistently across all environmental conditions, so we compared the Monte Carlo-derived rates for the kinetic Ising model with the averages of Monte Carlo, eigenvalue, and MFPT estimates for the 1D and 2D representations for a "physiologically relevant" range in values of $N$, $T$ and $V$. The results are shown in Fig. 10.

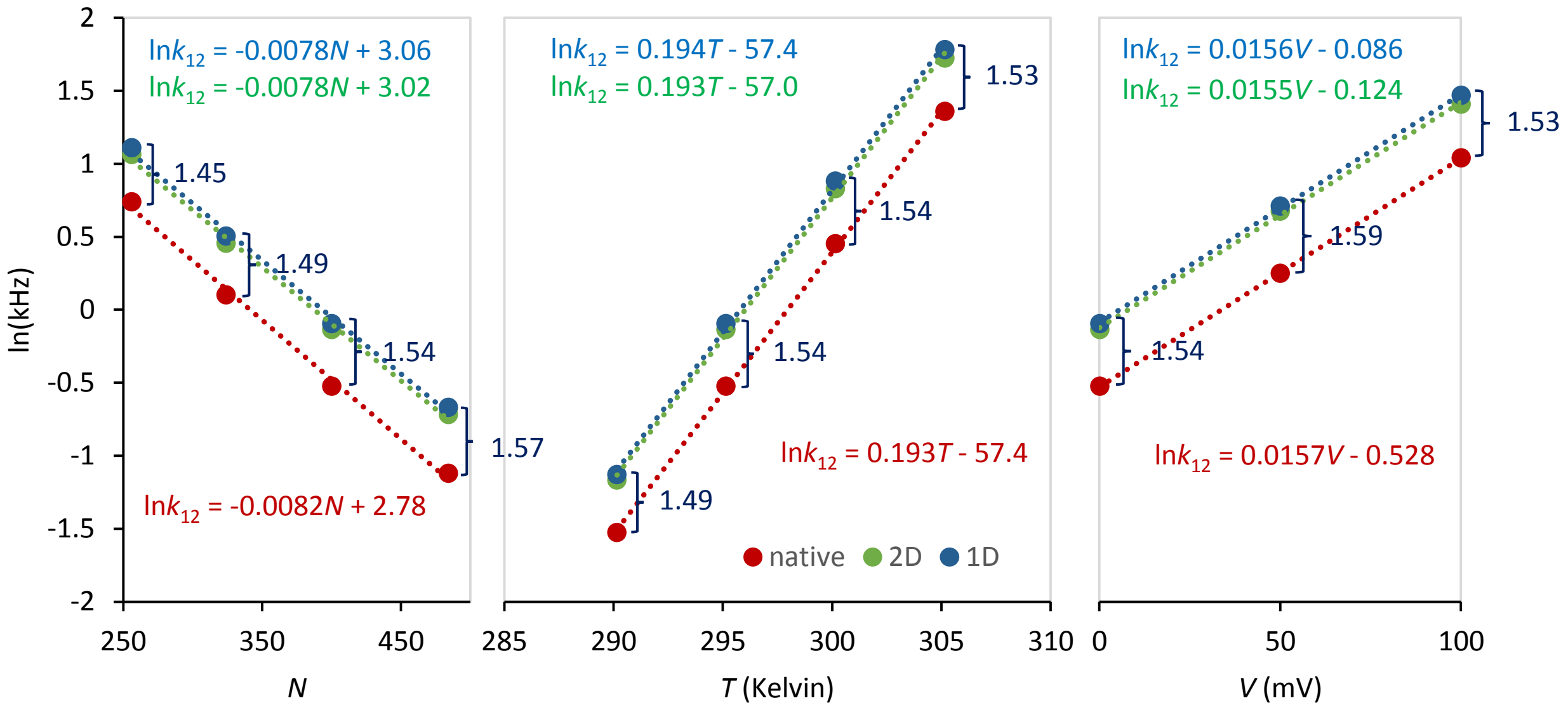

FIG. 10. Linear regression analysis in the semilog plot of forward rate constants with respect to the same environmental conditions as in Fig 9. The Ising rate constants were calculated as the inverse means of Monte Carlo first passage times from $q_1$ to $q_2$, as described in the text. The 1D and 2D rate constants were averaged from their respective Monte Carlo, MFPT, and eigenvalue calculations. Error bars from six experiments are not visible because they are located within the markers. The linear regression equations are color-matched to their respective fitted data. The ratios of 1D to Ising rates are written in dark blue.

The results show that for all conditions, the coarse-grained (1D and 2D) estimates of forward rate constants remained about 50% greater than the Ising rate constant, with 1D rates again slightly greater than their 2D counterparts. The approximate 3:2 ratio was consistent, though there was a small tendency towards larger ratios with increasing system size ($N$). Linear fits describing the relationships $\ln k_{12} = cX + d$, with $X$ representing either $N$, $T$, or $V$, resulted in nearly identical slopes $c$ for Ising, 1D, and 2D calculations. These slopes have physical significance. From the $\ln k_{12}$ vs. $N$ plots, one concludes that every added particle increases the free energy barrier by an average of 0.2 meV. The ln$k_{12}$ vs. $T$ and ln$k_{12}$ vs. $V$ plots are consistent with an activation charge of 0.4 eu and an activation energy of 1450 meV.

Our findings suggest that, within a "physiological" range of environmental variables, one could correct the coarse-grained rate constants by dividing by a constant factor of 1.5. The small difference between 1D- and 2D-derived values of $k_{12}$ can be attributed to the modest concavity of the 2D reaction pathway (Fig. 2). There is presumably some dynamic information lost when the 2D landscape is projected onto a 1D pmf. The 3:2 ratio in coarse-grained versus experimental rates is consistent with the same ratio for the diffusion coefficients $D_b$ and $D_{exp}$ (Fig. 5), as both quantities were calculated with the same pmf. We wanted to see if Nyquist analysis, which does not require coarse-graining, yielded a consistent ratio of $D_N$ to $D_b$ for different conditions. The means and standard deviations of $D_N/D_b$ for seven experiments varying $T$ and $N$ in Fig. 11 was 0.952 ± 0.004. The 30% simulation window used was optimal for obtaining the largest value of $D_N$. A smaller window lead to edge effects, and a larger window exceeded the linear region of $D(q)$. The constancy of $D_N/D_b$ suggests that $D_N$ closely follows the coarse-grained estimates of $D$.

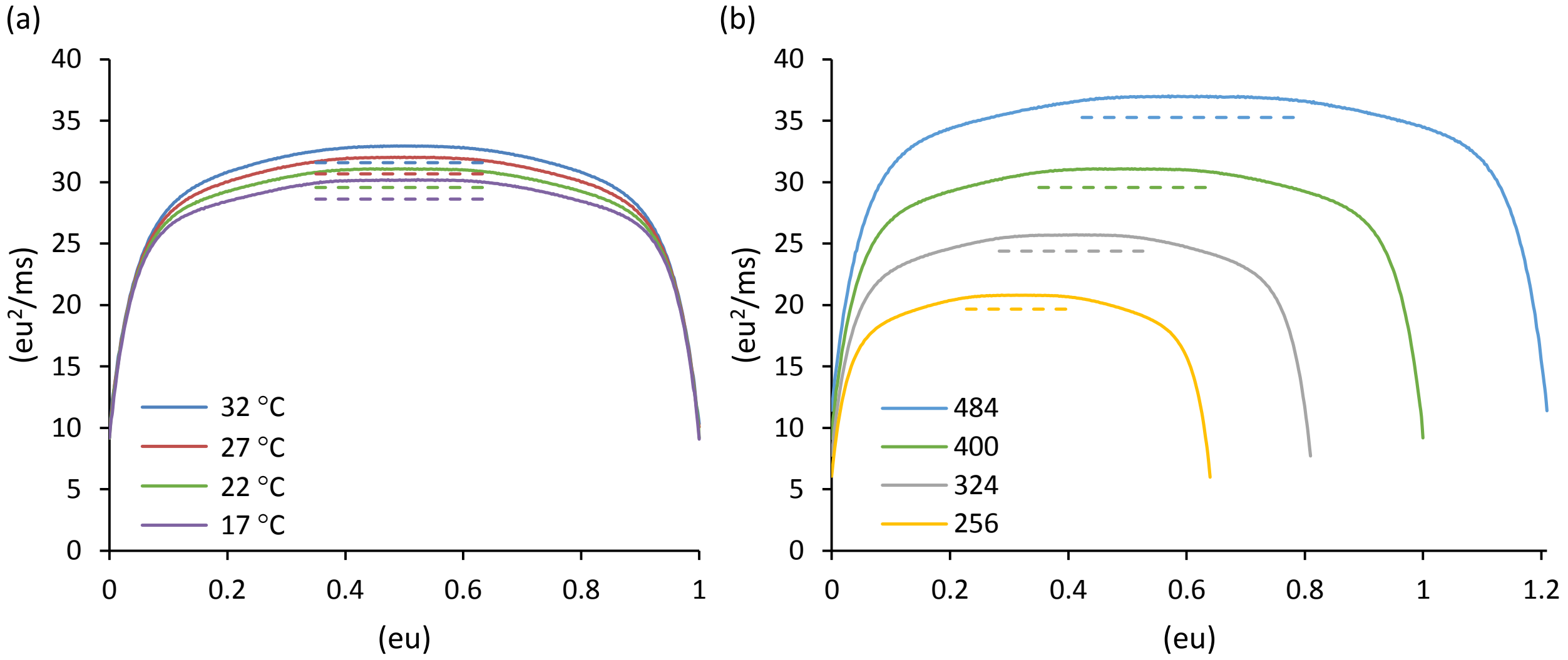

FIG. 11. Diffusion landscape $D(q)$ as a function of (a) $T$; (b) $N$. Dashed lines indicate the Nyquist value $D_N$ obtained from a 30% simulation window; $f_c = 5 \times 10^5$ kHz. The green curve in both plots is the reference model ($N$ = 400; $T$ = 22 °C; $V$ = 0 mV).

## V. DISCUSSION

The inspiration for this study came about after one of us (DS) attempted to construct a 1D diffusion landscape for the kinetic Ising model. After efforts to predict the correct Ising transition rate from the coarse-grained landscape failed after using several methods (Monte Carlo, Langevin, MFPT, and eigenvalue), it was initially thought there was a problem with the 1D projection along $q$, so we expanded the landscape to a 2D representation in $q$ and $E$ (microcanonical ensemble). The 2D model provided a complete thermodynamic solution to the finite Ising model, but yielded only a small improvement in the value of the rate constant. It became clear the main culprit was the manner in which microscopic rates were averaged at the mesoscopic level. What works for thermodynamics may not work for kinetics.

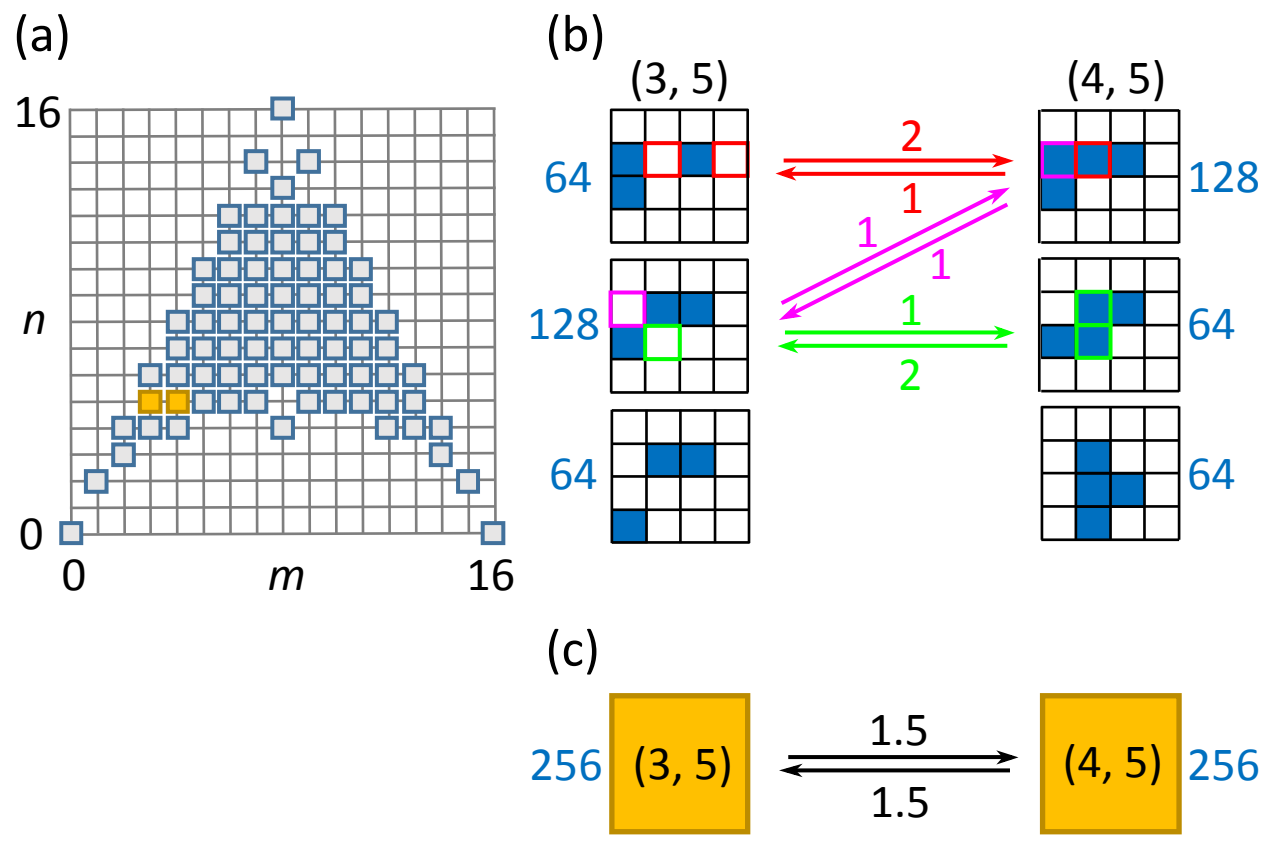

FIG. 12. (a) Mesoscopic ($m$, $n$) grid for the 4 x 4 Ising model. The (3, 5) and (4, 5) states are colored yellow. (b) Configurational sub-groups compatible with the (3, 5) and (4, 5) states. Blue cells are activated. The number of microstates ($\Omega_i$) in each sub-group are in blue. Allowed transitions between sub-groups are labeled with a local frequency factor that is color-coordinated to the cell(s) responsible for the transition. (c) The mesoscopic transition scheme, where the averaged frequency factors (= 1.5) are the $\langle r \rangle$ rates.

While our methods did not reveal the precise source of coarse-graining error, a clue may be found by analyzing in detail a specific transition in a small ($N$ = 16) Ising system (Fig. 12). In this 4 x 4 model, the mesostates (3, 5) and (4, 5), which are adjacent and have the same energy, both contain $\Omega$ = 256 microstates split into three configurational substates ($i$), with each substate containing $\Omega_i$ microstates. Two of the three substates allow transitions to the opposing mesostate. These local transitions contribute to the mean rate constant in proportion to the product of $\Omega_i/\Omega$ and a local frequency factor (the number of transition paths available to the sub-state). The average of these products is the mesoscopic frequency factor $\langle r \rangle$. In contrast to the first two substates, the third substate, comprising a quarter of microstates in the total mesostate, does not contribute to the reaction because there are no transition paths available to it. Should the system find itself in one of the microstates of this non-reacting group, a forward transition in $q$ must use a path to a higher or lower energy state. This suggests that for the Ising model, there are fewer trajectory pathways available than predicted by the coarse-grained representation. We conjecture that rate constant averaging over the microcanonical grid does not properly account for the smaller possible number of trajectories, and thereby overestimates the true rates.

It is not always feasible to coarse-grain a system into a smaller network of states for the purpose of solving kinetics, as we did here for the discrete-state Ising model using the microcanonical ensemble. In real-world applications involving proteins and other macromolecules, states are continuous and rate constants are not known or not well defined, although there has been much progress made in coarse-graining MD trajectories using the method of Markov state modeling [41–44]. An alternative approach to solving protein kinetics is to define a pmf with respect to a chosen set of collective variables, and by adding a diffusion coefficient, generate a Fokker-Planck or Langevin equation that can be used to calculate macroscopic rates. Van Kampen has termed this ad hoc method the "Langevin approach", and criticizes it for not properly reflecting the underlying mechanics of the system [45]. Nevertheless, such phenomenological models serve to contextualize experimental data, whether obtained from computer simulation or from the laboratory. For large-scale protein fluctuations such as the folding process or enzymatic reactions, where dynamics are assumed to be spatially diffusive, the nonlinear Smoluchowski equation has been used [15,27,46,47]. This requires knowing the (possibly inhomogeneous) diffusion coefficient, which can be derived from fluctuation analysis [31] or by comparing outcomes to experimental data [16]. In this paper, we used both methods. The Nyquist estimate $D_N$ was obtained from large-bandwidth fluctuations at the barrier peak, whereas $D_{exp}$ was obtained as an adjustable value by fitting the pmf-derived rate constant to the experimental rate constant. Since the diffusion coefficient at the barrier determines the reaction rate, it is valid to compare $D_N$ to $D_{exp}$, and also to $D_b$, the critical barrier value for the coarse-grained 1D landscape $D(q)$. The finding that the value of $D_N$ is much closer to $D_b$ than $D_{exp}$ is somewhat unexpected, since both $D_N$ and $D_{exp}$ were obtained from simulation of the Ising model prior to coarse-graining. Since the Nyquist formula is based on the fluctuation-dissipation theorem [48], the finding that $D_N \approx D_b$, with both quantities being substantially greater than $D_{exp}$ suggests that linear response does not hold for the kinetic Ising model, resulting in a breakdown of the fluctuation-dissipation theorem.

Given our results, it is conceivable that kinetic coarse-graining, either through the use of ensemble-averaged rate constants or constructing a diffusion landscape, will fail for any complex system with inter-particle interactions. In the case of the Ising model, it is encouraging that the approximate 3:2 ratio of calculated to simulated rates appears fairly constant across "physiological" conditions. Thus, for practical purposes, the two-state kinetics of our toy model has been "solved", provided one employs the correct scale factor. It remains to be seen whether such a scale factor exists for real-world systems.

## ACKNOWLEDGMENTS

We thank all the participants of the CECAM workshop "Ion Transport from Physics to Physiology: the Missing Rungs in the Ladder" for excellent discussions that inspired this investigation. This work was supported in part by the National Institute of General Medical Sciences of the National Institutes of Health under award numbers: R01GM093290 (V.C.); R01GM123296 (V.A.V.); S10OD020095 (V.C. and V.A.V.).This research was supported in part by the National Science Foundation through major research instrumentation grant number 1625061 and under award number ACI-1614804 (V.C.).